\documentclass[useAMS,usenatbib]{mn2e}

\title[Age distributions of clusters in galaxies]
{Age distributions of star clusters in spiral and
barred galaxies as a test for theories of spiral structure}
\author[C. L. Dobbs \& J. E. Pringle]
{C. L. Dobbs\thanks{E-mail:
dobbs@astro.ex.ac.uk}$^{1,2} $\& J. E. Pringle$^{3,4}$\\
$^1$ Max-Planck-Institut f\"ur extraterrestrische Physik, Giessenbachstra\ss{}e, D-85748 Garching, Germany \\
$^2$ Universit\"ats-Sternwarte M\"unchen, Scheinerstra\ss{}e 1, D-81679
M\"unchen, Germany\\
$^3$ Institute of Astronomy, Madingley Road, Cambridge, CB3 0HA \\
$^4$ STScI, 3700 San Martin Drive, Baltimore, MD~21218, USA \\}
\usepackage{amssymb}
\usepackage{amsmath}
\usepackage{graphicx}
\usepackage{epsfig}
\usepackage{multirow}

\begin{document}
\date{\today}

\pagerange{\pageref{firstpage}--\pageref{lastpage}} \pubyear{0000}

\maketitle

\label{firstpage}

\begin{abstract}

We consider models of gas flow in spiral galaxies in which the spiral
structure has been excited by various possible mechanisms: a global
steady density wave, self-gravity of the
stellar disc and an external tidal interaction, as well as the case of
a galaxy with a central rotating bar. In each model we
estimate in a simple manner the likely current positions of star
clusters of a variety of ages, ranging from $\sim 2$~Myr to around
$\sim 130$~Myr, depending on the model. We find that the spatial
distribution of cluster of different ages varies markedly depending on
the model, and propose that observations of the locations of age-dated
stellar clusters is a possible discriminant between excitation
mechanisms for spiral structure in an individual galaxy.

\end{abstract}

\section{Introduction}

Understanding the nature and origin of spiral structure in galaxies is
still a fundamental problem in astrophysics. It is evident that in
many galaxies spiral structure is driven by tidal effects, either
externally by interaction with a companion, or companions, or
internally by a central rotating bar. It is worth noting than the
possibility of tidal excitation is not precluded by the apparent
absence of a visible perturber, because orbiting clumps of dark matter
might be responsible \citep{Tutukov2006,Dubinski2008}. Indeed, some authors
have argued that all spiral structure is driven in this manner
(e.g. \citealt{Kormendy1979,Bottema2003}). However determining whether
{\it all} spiral structure is tidally induced is not
a simple observational problem.

There is also the possibility, at least from theoretical
considerations, that some spirals can be self-excited. Here two
distinct possible mechanisms for the excitation of spiral structure
have been proposed. First, there is the argument that spiral structure
is the result of quasi-steady, global modes in the stellar disc
\citep{Lin1964,Bertin1989}. 
This is commonly referred to as
`density wave theory'. Second there is the suggestion that spiral
structure is the result of local, transient self-gravitational
instabilities \citep{Toomre1964} which shear to form short-lived spiral
arms \citep{Toomre1990,Sellwood1991}. Early numerical simulations of this process
\citep{Sellwood1984} indicated that such activity might be
short lived, in the absence of some other phenomenon, such as gas
accretion, which could act to reduce the stellar dispersion velocity,
and so maintain the disc close to the borderline for
self-gravitational instability.  However, more recent, higher
resolution simulations suggest that the activity can last for a
substantial fraction of a Hubble time \citep{Fujii2010}.

One problem, highlighted in a recent review \citep{Sellwood2010}, is
that it is difficult to find observational tests to distinguish
between the various theoretical models. In fact the usual
observational approach seems to be to assume one model (usually the
assumption is that there is a well-defined global pattern speed), and
to then interpret the observations in terms of that particular model
(e.g.  \citealt{Elmegreen1989c, Kendall2008,
Mart2009}).  Several authors have tried to infer
pattern speeds by use of colour gradients to observe a transition in
stellar ages across the spiral arms
\citep{Efremov1982,Regan1993,Beckman1990,Gonzalez1996,Mart2009}. However
with the exception of one or two cases, there is rarely an
identifiable trend in the colour gradients.

For nearby galaxies, it is also possible to study the ages of
individual clusters
\citep{Anders2004,Fall2005,Scheepmaker2007,Fall2009, Bastian2009}. 
This technique involves fitting the \textit{UBVI} H$\alpha$ magnitudes
to those predicted by stellar population and starburst models, and
then estimating an age for each star cluster. Estimates of ages are
thought to be accurate to within factors of around two, and ages can be
derived from around 1~Myr to around 100~Myr. This method has been
applied to the LMC and SMC
\citep{Hunter2003,deGrijs2006,Chandar2010}, and interacting
galaxies, the Antennae \citep{Fall2009} and M51
\citep{Bik2003,Kaleida2010}. 
So far, the age-dating of clusters has been largely used to examine intrinsic
properties of clusters, e.g. the cluster IMF, age mass relation, size
age relation. However if one can
assume that in spiral galaxies such stellar clusters form
predominantly in regions of high gas density, for example
predominantly within the spiral arms, then the distribution of
age-dated clusters throughout the spiral galaxy can be used as a means
of obtaining some idea of the nature and timing of the gas flow
relative to the spiral arms. In this way it might be possible to
distinguish between the various theoretical models for spiral arm
formation.

\citet{Bash1977} also made predictions about stellar 
  clusters in the context of density wave
  theory. However their analysis concentrated on the observational
  charcteristics related to ongoing star formation, e.g. the presence
of O stars, CO emission. Furthermore, the birth sites of their
clusters were assumed to lie along an imposed spiral, rather than
derived from numerical models.

In this paper we carry out numerical simulations of gas flow in model
galaxies in which the spiral arms are excited by the various
theoretical mechanisms and then locate the regions in which `star
clusters', formed in the dense arm regions, can be expected to be
found at later times. We consider four canonical galaxy models,
corresponding to the four basic theoretical models for spiral arm and/or bar
formation. These are: (i) a galaxy with an imposed spiral potential
with fixed pattern speed, (ii) a galaxy which is bar unstable, (iii) a
`flocculent' galaxy in which the arms are intermittent and driven by
local gravitational instabilities, and (iv) a galaxy which is subject
to a strong external tidal interaction.  We then make predictions for
the distribution of stellar clusters of different ages. We find that
different spatial distributions arise at different cluster ages
depending on the underlying dynamics of the galaxy, and on the spiral
excitation mechanism. We finally suggest that by using methods for 
age-dating clusters, e.g. in the recent work by \citet{Fall2009}, it may be possible
to identify what is the underlying mechanism producing spiral
structure in individual nearby galaxies.

In Section~\ref{models}, we outline our basic numerical model, both
for the galaxy dynamics, and for the identification of locations for
age-dated star clusters. We then apply these techniques to the four
excitation mechanisms mentioned above: `density wave theory' in
Section~\ref{linshu}, bar-driven waves in Section~\ref{bar},
flocculent behaviour in Section~\ref{flocculent} and external tides in
Section~\ref{m51}. In Section~\ref{xsection} we illustrate our
findings by plotting the distributions of cluster ages across spiral
arms for the different excitation mechanisms. We summarise our
findings and provide discussion in Section~\ref{discussion}.

\section{ Simulations }
\label{models}

All the calculations presented here use an SPH code, developed
originally by \citet{Benz1990} and substantially modified to include
sink particles \citep{Batesph1995}, individual timesteps
\citep{Bate1995} and individual smoothing lengths \citep{Price2004}.

For all the calculations except the fixed spiral
(Section~\ref{linshu}), we fully model the galaxy with particles for
stars, gas, and the dark matter halo. Since we are primarily
interested in the gas flow, and in particular the gas flow downstream
of any density maxima which might be taken to give rise to star
cluster formation, we use relatively low gas surface densities. In the
case of the fixed spiral we do not include the self-gravity of the
gas. In the models which do include gas self-gravity
(Sections~\ref{bar}, \ref{flocculent}, and~\ref{m51}) we choose a
relatively high gas temperature ($10^4$ K), in order to avoid
widespread gravitational collapse. In all models the gas is taken to
be isothermal. Thus we make no pretence of modelling star-formation,
feedback, radiative processes in the ISM, and so on, in any
detail. Rather we are trying to identify those regions in which the
gas tends to have higher than average density, and will then identify
those regions as the ones in which star clusters are most likely to
form. Even so, in the runs with self-gravity, we have still had to
insert a few sink particles to replace the highest density regions in
order to ensure that the simulations did not take too long. The number
of sink particles used is low (see
Sections~\ref{bar}, \ref{flocculent} and \citealt{Dobbs2010}).

\subsection{Estimating cluster ages}
\label{ages}

Since in our simulations we cannot resolve the formation of star
clusters we instead locate dense gas, which would be where stars are
more likely to form, given a more realistic surface density (or
self-gravity, for the fixed spiral) and sufficient resolution. For
each simulation, we locate gas at high density at times between 2~Myr
and 130~Myr before a given time frame. So for example, for the
calculation with a fixed spiral potential (Section~\ref{linshu}), we
take a time frame of 255~Myr to represent the present. We then locate
dense gas at times of 125~Myr, 155~Myr, 205~Myr and 253~Myr.  We
assume that the dense gas represents the location of stellar clusters
forming at those times, and then plot the location of that gas at the
time of 255~Myr. Thus we obtain estimates for the location of star
clusters of ages of 2~Myr, 50~Myr, 100~Myr and 130~Myr. The definition
of `dense gas' requires some care, since in general the mean gas
density decreases with radius. In practice, we found for the models
except the fixed spiral, it was sufficient to
select gas which has a density more than 3 times the average (mass
weighted) surface density for a given radial bin, as being regions where star
formation would be likely to take place. For the fixed spiral (Section~3), we used
colder gas, which shocks to higher densities, and thus chose a density
of 10 times the average. Radial bins were chosen to
have width $\Delta R/R \approx 0.08$.

As we do not explicitly include star formation in these simulations,
we cannot distinguish between SPH particles which represent star
clusters and those which represent gas.  Thus, we assume that the
trajectories of the gas particles are not dissimilar to stars. In
reality stellar clusters lose their gas over a time period of a few
Myr, after which they are not subject to gas pressure. However, the
role of gas pressure is most important as gas passes through a spiral
shock. In the simulations, the dense gas that we assume represents the
locations of star cluster formation has already passed through the
shock. The velocity of the gas is highly supersonic and thus the
effect of gas pressure is in general small, except in shocks. We
therefore expect those stars formed in an arm, together with the gas
from which they form, to emerge from a spiral arm with similar space
velocities, and to continue on neighbouring trajectories until they
pass through the next spiral shock. In our calculations, the majority
of the cluster locations we identify occur
before the gas has gone through the next spiral arm. We test this
assumption that gas pressure is, for the most part, negligible,
explicitly in Section~3. 
 

\section{Fixed global pattern speed -- `density wave theory'}
\label{linshu}

The model galaxy discussed in this Section is subjected to a fixed
spiral potential rotating with a fixed global pattern speed.  It is
intended to represent the \citet{Lin1964} model of quasi-steady spiral
structure, which predicts the presence of a global spiral mode. 
We note that a galaxy which exhibits a
simple kinematic density wave (\citealt{Binney1987}, Section 6.2;
\citealt{Dobbs2010}, section 6.1.2) which has a radially-dependent
pattern speed will show analogous effects. We
perform a calculation analogous those presented in
\citet{Dobbs2006,DBP2006}. We impose a fixed cylindrically symmetric
`galaxy' potential, comprising of a logarithmic component, which
produces a flat rotation curve with a peak velocity of 220 km
s$^{-1}$, and add an $m=4$ spiral perturbation from
\citet{Cox2002}. Thus unlike the other galaxy models presented below,
we only need to model the gas component of the disc.  The spiral has a
fixed pattern speed and has an amplitude of about 4 per cent (see
\citet{Dobbs2008}).  The corotation radius is at $R_{\rm c} = 11$ kpc.
For this calculation, there is no self-gravity, and the gas is
taken to be isothermal with a temperature of 100 K. We use 1 million
gas particles, which are initially distributed randomly in the disc,
so that the gas has an initial uniform surface density.

The simulation was run for a total of 450 Myr, and the rotation period
at the midpoint of the torus ($r=7.5$ kpc) is 200 Myr. 
During the course of the simulations, the gas becomes compressed 
into clumps in the spiral arms, which are then sheared into spurs
\citep{DBP2006}. A PDF of the density
distribution of the gas remains roughly constant after about 125
Myr. The morphology of the spurs still varies with time, but as they
do not greatly change the gas flow, this is not
likely to effect our results.

Our method for estimating cluster ages is the simplest for the fixed
pattern speed scenario, since, after initial transient behaviour, the
dynamics is essentially stationary in the frame rotating with the
pattern speed. In the rotating frame, we expect the gas to follow steady
streamlines. The simple expectation is that older clusters appear at
correspondingly larger distances downstream once the gas has passed
through the spiral arms. The domain of our simulation lies entirely
within corotation, and thus the older stars are always expected to be
the same side of the spiral arms.

In Fig.~1a we show the surface density of the gas at a representative
time of 225~Myr.  In (Fig.~2a), we plot the positions of notional star
clusters which formed at times 2~Myr, 50~Myr, 100~Myr and 130~Myr
earlier than this time.  We see that the youngest star clusters are
all found in the spiral arms. The gas actually spends relatively long
(around 60~Myr) in the spiral arms \citep{DBP2006}. Hence even after
50 Myr, star clusters lie still predominantly in the spiral arms. But
after 100 Myr, the stars have moved into the interarm region. After
130 Myr, the stars have moved further into the interarm region and are
approaching the next spiral arm. Thus we observe a continuous and
monotonic increase in the ages of star clusters across the spiral
arm. This is what one would expect from standard theory
\citep{Fujimoto1968,Roberts1969,Roberts1972}.  To
illustrate this more clearly, in Figure~3 we plot the paths of
representative particles for this model in the rotating frame of the
potential. If the flow were exactly steady, these paths would
represent gas streamlines, but as can be seen there is some mixing
present in the spiral shock. We mark points corresponding to the
predicted locations of stellar clusters with ages 2, 50 and 100
Myr. These points correspond to the locations in the particle's
trajectory at 2, 50 and 100 Myr after the first density maximum of the
gas in the shock (typically there is only one density maximum over the
duration the gas spends in the spiral arm).

We tested whether our assumption that gas pressure does not
  strongly change the locations of the clusters by running the
calculation with a fixed spiral potential from 205~Myr to 255~Myr, and
from 125~Myr to 255~Myr without gas pressure. The first case
corresponds to clusters of age 50 Myr, and the second 130 Myr. For the 50
Myr clusters, the mean difference in the locations of clusters with
and without gas pressure was 30 pc, and only a tiny fraction (2\%) of
the clusters are displaced by more than 100 pc. 
For the 130 Myr clusters, a
larger fraction (25\%) exhibit displacements of more than
100 pc, as the clusters go through the
next spiral arm. However 100 pc is still small compared to the
length scales in our simulations, so the overall distribution of
clusters calculated without pressure occupies a similar area to that 
shown in Figure~2, even after 130 Myr.

{We did not include self gravity in this calculation, which may
  alter the location and dynamics of the spiral shock
  \citep{Lubow1986}. However in view of observational evidence that gas
  clouds are actually slightly unbound
  \citep{Heyer2009,Koda2009}, the self gravity of the gas is likely to
  be much less important than assumed in the calculations of
  \citet{Lubow1986}.

Young star clusters are found in the so-called `spurs' (or feathers)
of spiral galaxies, downstream of the spiral shock. Spurs are the
result of large agglomerations of dense cold gas (typically Giant
Molecular Associations) which form in the spiral shock being sheared
out as they leave the arm in the downstream flow 
\citep{Dobbs2006,Koda2009,Muraoka2009}. However,
while spur formation is present in the current simulations (Figure~1),
the interarm densities recorded here are insufficient to be recorded
as sites of cluster formation. Thus there are no young star clusters
predicted in the spurs seen in Figure~2a. We note too that, while in
observed spirals some star formation does occur in interarm regions,
the bulk of the star formation does occur in spiral arms
\citep{Elmegreen1986}. 
Modelling such behaviour needs a more complex treatment 
of the ISM than we assume here (see, for example,
\citealt{Dobbs2009}), 
or it could be inserted {\it deo ex machina}
(e.g. \citealt{Karl2010}). Thus while the simpler models here enable us
to estimate where {\it most} of the star clusters are likely to form,
they cannot form the basis of a prediction of where {\it all} such
clusters might form.
\section{Barred galaxy}
\label{bar}

For this galaxy model, and indeed for the other two simulations
(Sections~\ref{flocculent} and~\ref{m51}), we model the whole galaxy
as a dynamical entity. That is, we include particles representing the
gas, the stars and the dark matter halo. For each calculation, the
initial conditions are generated using the mkkd95 program
\citep{Kuijken1995}.

To produce a barred galaxy, we simply set up a galaxy with a high disc
to halo mass ratio, since galaxies with a large fraction of their mass
in the disc are known to be unstable to $m=2$ perturbations
\citep{Ostriker1973,Kalnajs1974,Hohl1976}.  For our calculation, the
masses of the disc and halo are $2.9 \times 10^{10}$ and $3.6 \times
10^{10}$ M$_{\odot}$ respectively. Thus the disc mass is 80 per cent
that of the halo. We also include a bulge, with a mass of $7.6 \times
10^{9}$ M$_{\odot}$. The number of particles in the halo, stellar disc
and bulge are $10^5$, $10^5$ and $2 \times 10^4$ respectively, and
there are $9 \times 10^5$ gas particles.  The gas constitutes 1 per
cent of the mass of the disc, with a surface density of 3 M $_{\odot}$
pc$^{-2}$ and temperature $10^4$ K. The gas surface density initially
follows that of the stars, though settles into equilibrium within a
short timescale (see \citet{Dobbs2010})

The disc quickly becomes unstable and produces an $m=2$ spiral in the
stars and gas, which materialises into a bar after a time of around
450 Myr. The bar rotates with a relatively constant pattern speed of
38 km s$^{-1}$ kpc$^{-1}$ for the duration of the simulation (800
Myr), so that the co-rotation radius is at about $R_{\rm c} \approx
5.3$ kpc. An inner ring also forms at the centre
of the galaxy, with a radius of about 1 kpc.  Though we chose a low
surface density and high temperature, some regions still become
gravitationally unstable, and the inclusion of a few sink particles
was necessary to enable the computation to proceed. However only 8
such particles formed during the simulation and thus they do not
affect the dynamics. The gas surface density, and the location of the
sink particles, are shown at a representative time of 552~Myr in
Fig.~1b.

For this barred galaxy (Fig.~2b) we show the positions of `star
clusters' of notional ages 4~Myr, 10~Myr, 50~Myr and 100~Myr.  We
again see that the youngest star clusters are found to lie in the bar,
whilst the older star clusters lie progressively further from the bar
(see also the simulations by \citet{Wozniak2007}) on the trailing
side. Thus there is a similar pattern to the case with a fixed
spiral. This is not so surprising since the stellar distribution
changes little during the course of the simulation, so the pattern
speed is effectively constant. The youngest stars occur on the leading
side of the bar, after the gas has passed the minimum of the potential
associated with the stars. This offset from the stellar bar is
predicted from analysis with a fixed bar potential \citep{Roberts1979}
and is common in observations \citep{Sheth2002}.

In the inner ring, we find clusters of all ages. There is no obvious
trend in ages with position, and clusters of 50~Myr, 100~Myr and even
10~Myr old stars appear to be spread throughout the ring. This is
presumably because the dynamical timescale there is short (around 30
Myr). \citet{Sandstrom2010} recently studied the central
ring of NGC 1097, and find that emission in tracers other than CO is
widespread throughout the ring, with no apparent age transition, in
agreement with these results.

\section{Flocculent galaxy}
\label{flocculent}

The simulations of the flocculent spiral, and of the tidally induced
spiral (Section~\ref{m51}) are both described in more detail in
\citet{Dobbs2010}. We use the same initial galaxy model in each case,
except that for the tidally induced spiral, we also include an
orbiting perturber. We again set up the initial conditions using the
mkkd95 program to assign particles to the disc, halo and bulge. This
time the disc mass is chosen to be approximately 40 per cent of the
halo. Thus the galaxy is more stable to bar formation, and does not
form a bar at least over the duration of the calculation (600~Myr).
Similar to the barred galaxy (Section~\ref{bar}), the gas contains 1\%
of the mass of the disc, and is isothermal with a temperature of
$10^4$ K.

In the case of the flocculent galaxy, the galaxy exhibits numerous, but transient
spiral arms, with no grand design spiral pattern. Gravitational
instabilities in the stars lead to transient, local, short spiral arms
which are co-spatial both in the stars and in the gas. The distribution
of gas surface density at a representative time of 498~Myr is shown in
Fig.~1.

In Fig.~2c we show the positions of `star clusters' of nominal ages
4~Myr, 11~Myr, 50~Myr and 100~Myr. In a flocculent galaxy such as
this, there is no fixed spiral pattern; rather the spiral arms form
and dissolve in an intermittent and random fashion
\citep{Dobbs2007}. What we see in this case is that there is no
obvious or systematic change in cluster ages {\it across} spiral
arms. Rather there are many short sections of spiral arms, which have
stars of roughly the same age {\it along the arm}. This comes about
because of the mechanism by which spiral arms form in this case. The
arms form because of local, small-scale gravitational
instabilities. The gas falls into the so formed local potential minima
associated with the stars. At the same time the local minimum is
subject to the shearing motion of the galaxy flow as a whole. The gas
is drawn out to form a local spiral arm segment, with the density
changing more or less simultaneously at all positions along the
segment. Thus the stars that form in the arm segment do so more or
less at the same time. As time proceeds, the spiral arm dissolves,
with the maxima in gas and stars being sheared out. This has the
result that in the simulation, the spiral arms, or arm segments,
typically contain either the young clusters (4~Myr or 11~Myr), or the
older clusters which are 50~Myr, or 100~Myr old (but generally not
both). However the distribution of cluster ages in some spiral arms is
somewhat more complicated, due to collisions or mergers between arms of different
ages. Furthermore the timescales are shorter at smaller radii, and in
the central regions, gas from one spiral arm can end up in a
different spiral arm by 100~Myr.

\section{Interacting galaxy}
\label{m51}

In our final example, we present results for a representative model of
spiral structure induced by tidally interactions. This particular
calculation was designed to model M~51, and is described in detail in
\citet{Dobbs2010}. The galaxy is set up in the same way as the
flocculent model, but is then placed on an orbit with a perturbing
galaxy of 1/3 the mass. The orbits of the two galaxies are taken from
\citet{Theis2003}, who performed N-body calculations designed to
reproduce M~51, and the initial locations and velocities of the
galaxies are stated in \citet{Dobbs2010}. The perturbing galaxy
NGC~5195 is represented by a point mass in our calculations.

The tidal perturbation in the interacting case leads to a grand design
$m=2$ spiral structure (Fig.~1). The nominal time of 301~Myr
represents the closest the simulation got to modelling the current
picture of M~51 (or NGC~5194) and its companion NGC~5195. At the time
shown in Fig.~1, the perturbing galaxy is actually passing close to the main
galaxy for the second time. In \citet{Dobbs2010} we show that unlike
the models with the fixed potential, the pattern speed of the spiral
is not constant, and that, rather, the pattern speed decreases with
radius. We also found further departures from simple density wave
theory, such as bifurcations and kinks along the spiral arms. Many such
features are caused because the perturbing galaxy actually undergoes
two close passages with M~51.

If the dynamics of the tidally induced spiral were to follow standard
density wave theory, as a global mode with a radially independent
pattern speed, or as a simple kinematic density wave, 
we would expect a similar scenario to Fig.~2a,
where there is a monotonic transition of ages across each spiral
arm. Instead, as can be seen in Fig.~2d, the distribution of cluster
ages (here 3~Myr, 10~Myr, 50~Myr and 100~Myr) is much more
complicated.  At some locations (such as Point A in Fig.~2d), we can
see cluster ages increasing across the arm in a manner similar to what
one might expect from standard density wave theory. However, what is
remarkable here is that while the 3~Myr clusters are on the inner edge
of the arm, and the 50~Myr clusters are on the outer edge of the arm,
there are no intermediate age 10~Myr clusters in between them. At
other locations (for example Point B) we see younger clusters entering
the spiral arm which already consists of older (50~Myr) clusters. And
in contrast, at Point C, we see younger (10~Myr) clusters moving out
of the spiral arm, whilst the older (50~Myr) clusters still lie in the
spiral arm. Generally there is a tendency for clusters of quite
different ages to be in the same spiral arm, but the ordering of ages
of clusters across the spiral arm can be in either direction. This
comes about because of the complicated dynamics of the
interaction. At large radii this is not surprising, since the arms
are largely tidal, with little gas flow through the spiral arm. But
even at lower radii, the distribution is much more messy compared with
the other cases.

Thus the difference in the distribution of clusters of a tidally
interacting spiral galaxy, compared to the case with a fixed pattern
speed (Section~\ref{linshu}) is evidently due to the much more complex
internal dynamics within the galaxy induced by the tidal
interaction(s) with the companion. In the interaction modelled here,
the companion passes close to the main galaxy twice, effectively inducing new
spiral arms on two different occasions. These spiral arms are then
able to evolve and interact.  This accounts for the asymmetry of the
spiral arms, large-scale branches, and the kink along one spiral arm
at $x\sim7$ kpc, close to Point A.

\begin{figure*}
\centerline{
\includegraphics[scale=0.28]{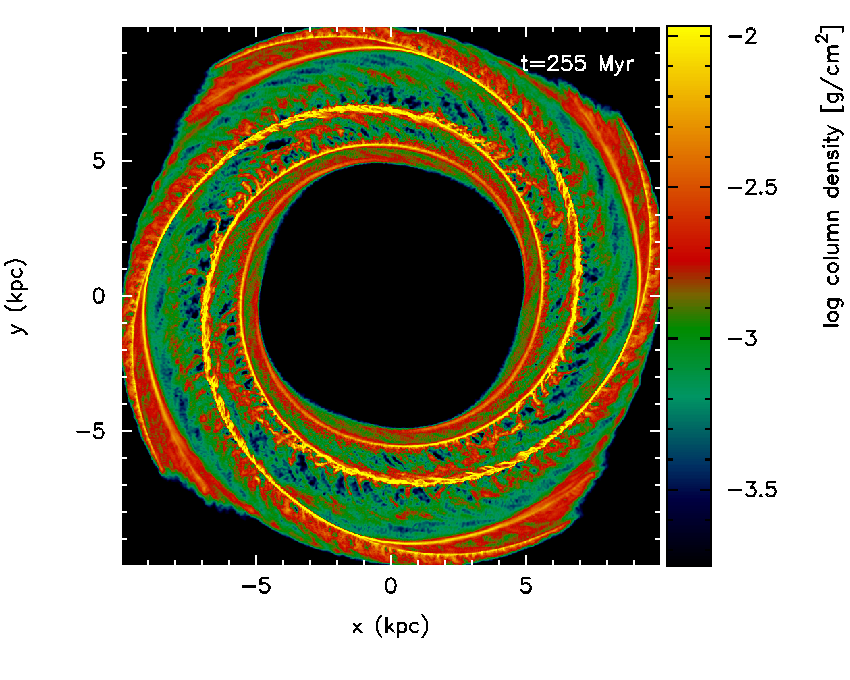}
\includegraphics[scale=0.28]{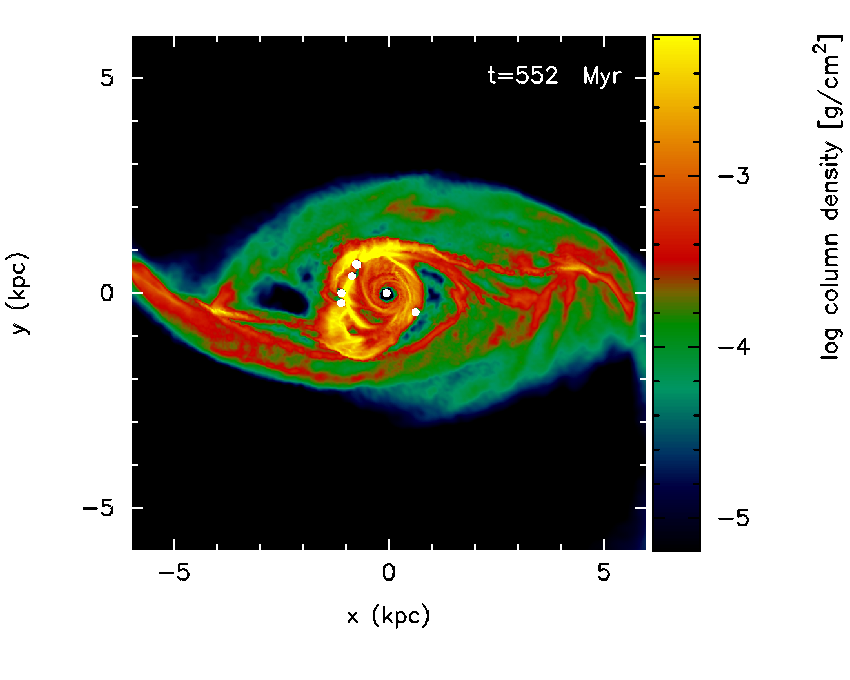}}
\centerline{
\includegraphics[scale=0.28]{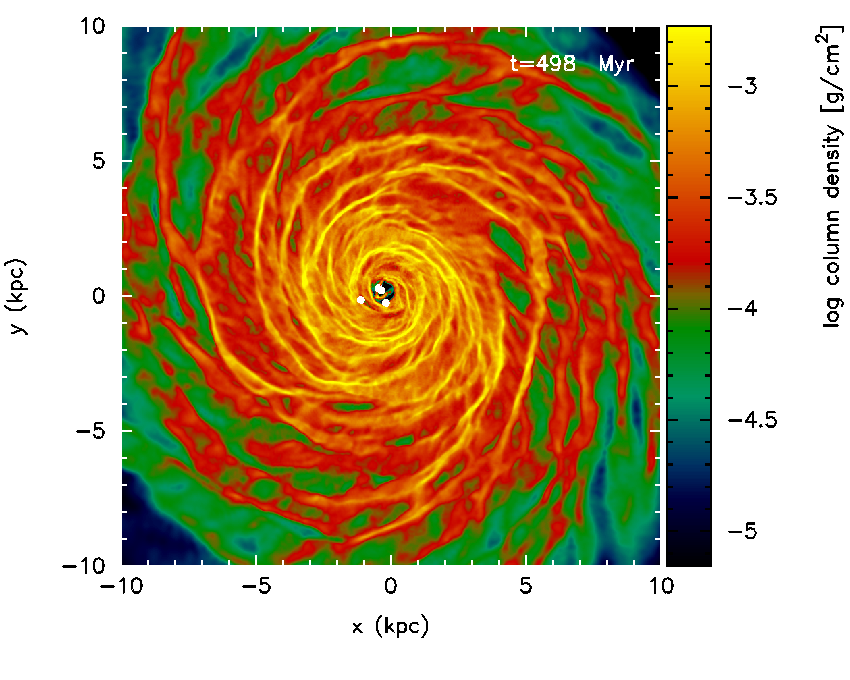}
\includegraphics[scale=0.28]{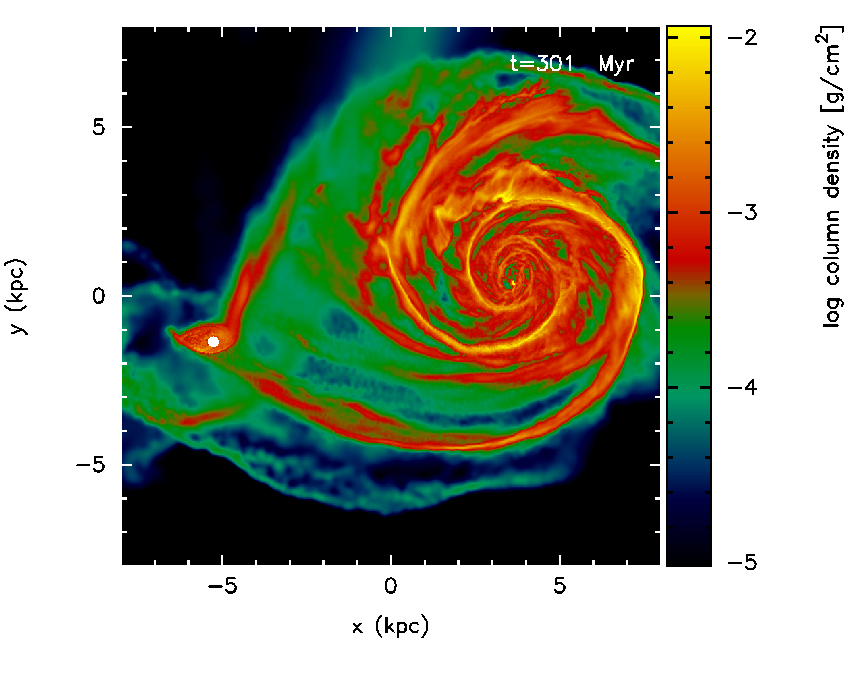}}
\caption{We display the gas column density for simulations of galaxies
  subject to various spiral excitation mechanisms. These mechanisms
  correspond to: (a) an imposed fixed spiral potential, representing a
  global mode from standard 'density wave theory' (top left), (b) a
  galaxy which has developed a central steadily rotating bar (top
  right), (c) a galaxy in which the stellar disc is unstable to
  self-gravity as a model of a flocculent spiral galaxy (lower left),
  and (d) and a galaxy subject to a tidal interaction (lower right).
  We show the expected locations of star clusters of different ages for
  these galaxies, at these times, in Fig.~2. For the barred and
  flocculent galaxies, the white dots represent sink particles. For
  the tidal spiral, the white dot shows the perturbing galaxy.}
\end{figure*}

\begin{figure*}
\centerline{
\includegraphics[scale=0.37]{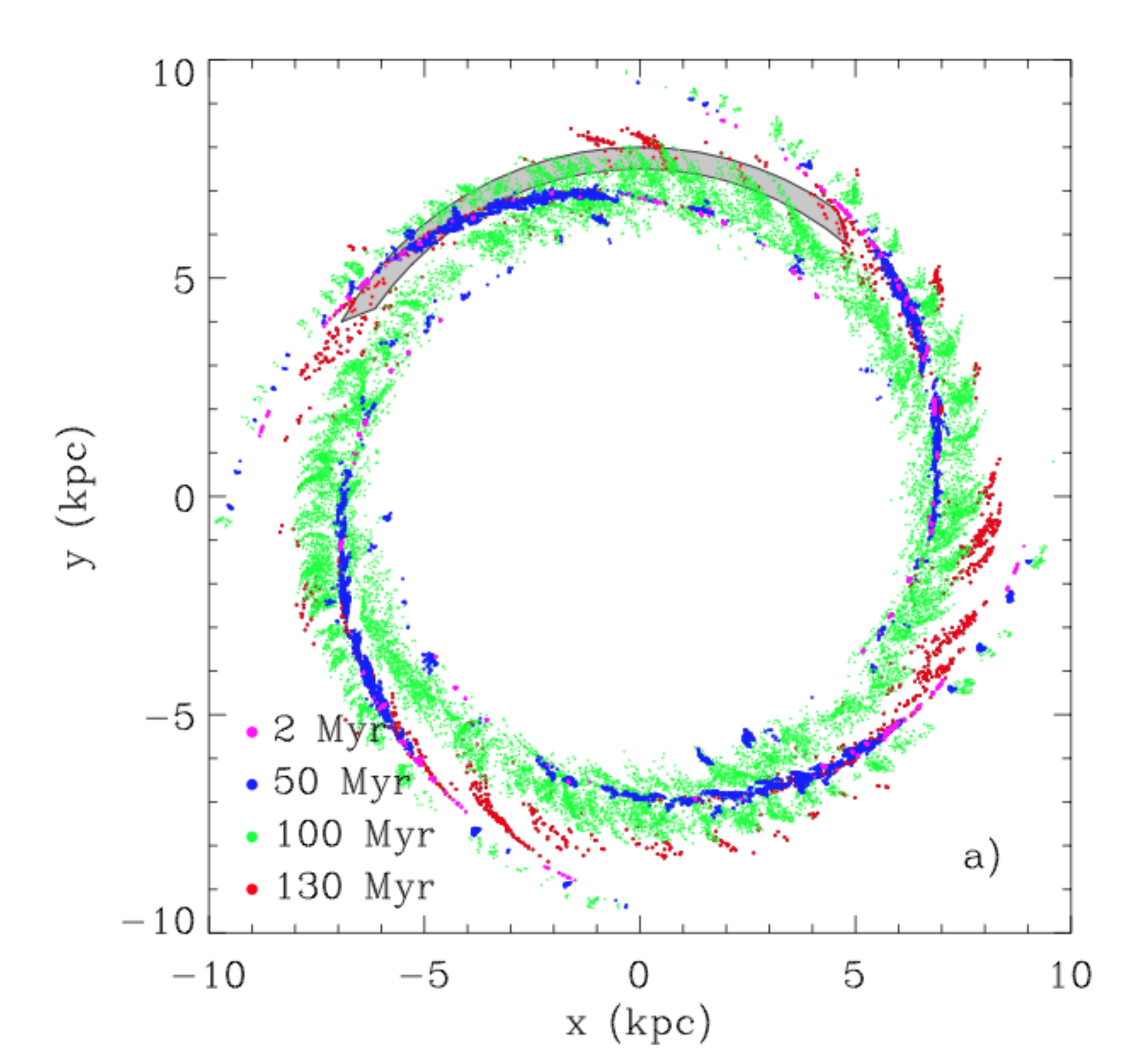}
\includegraphics[scale=0.37]{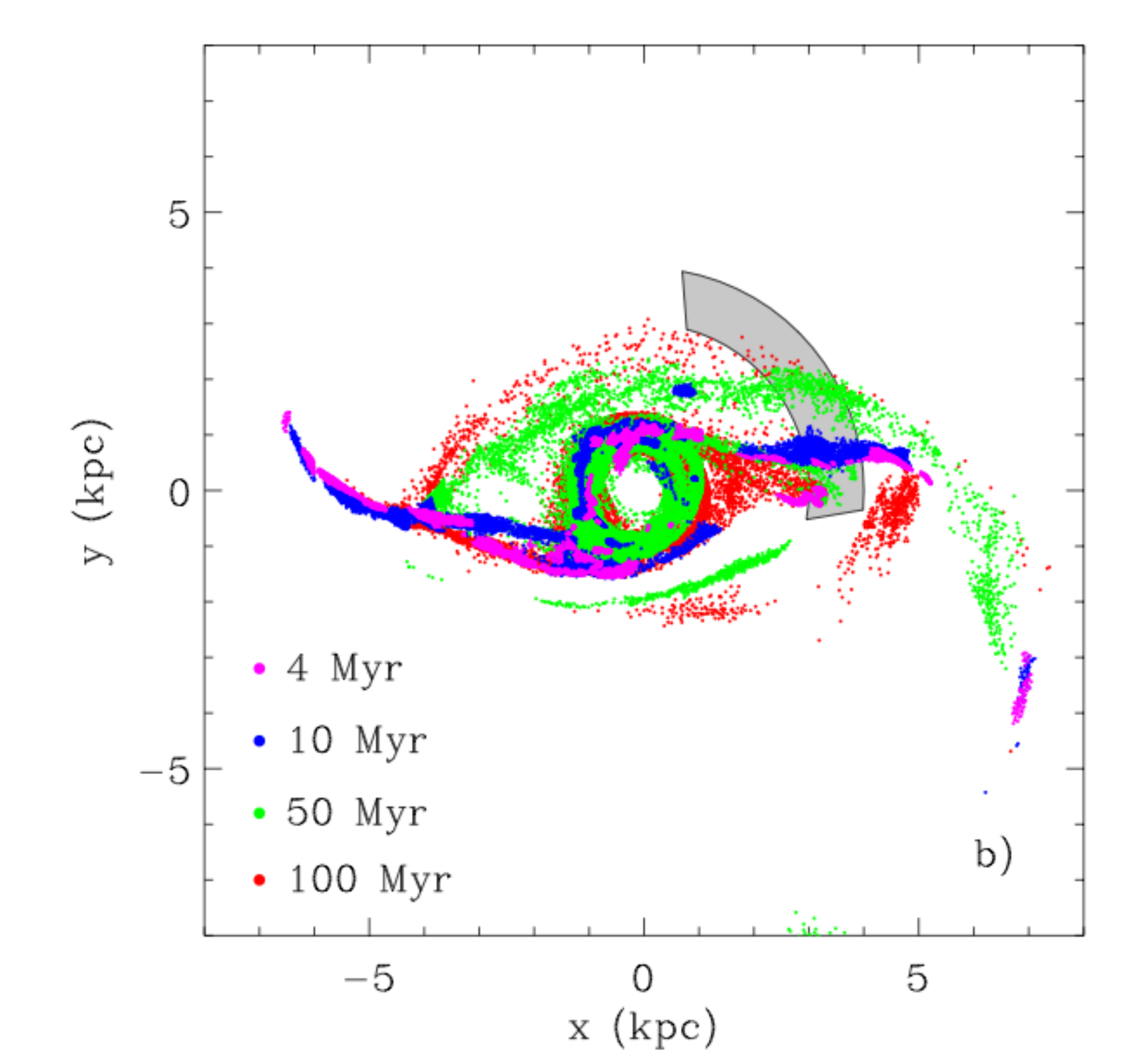}}
\centerline{
\includegraphics[scale=0.37]{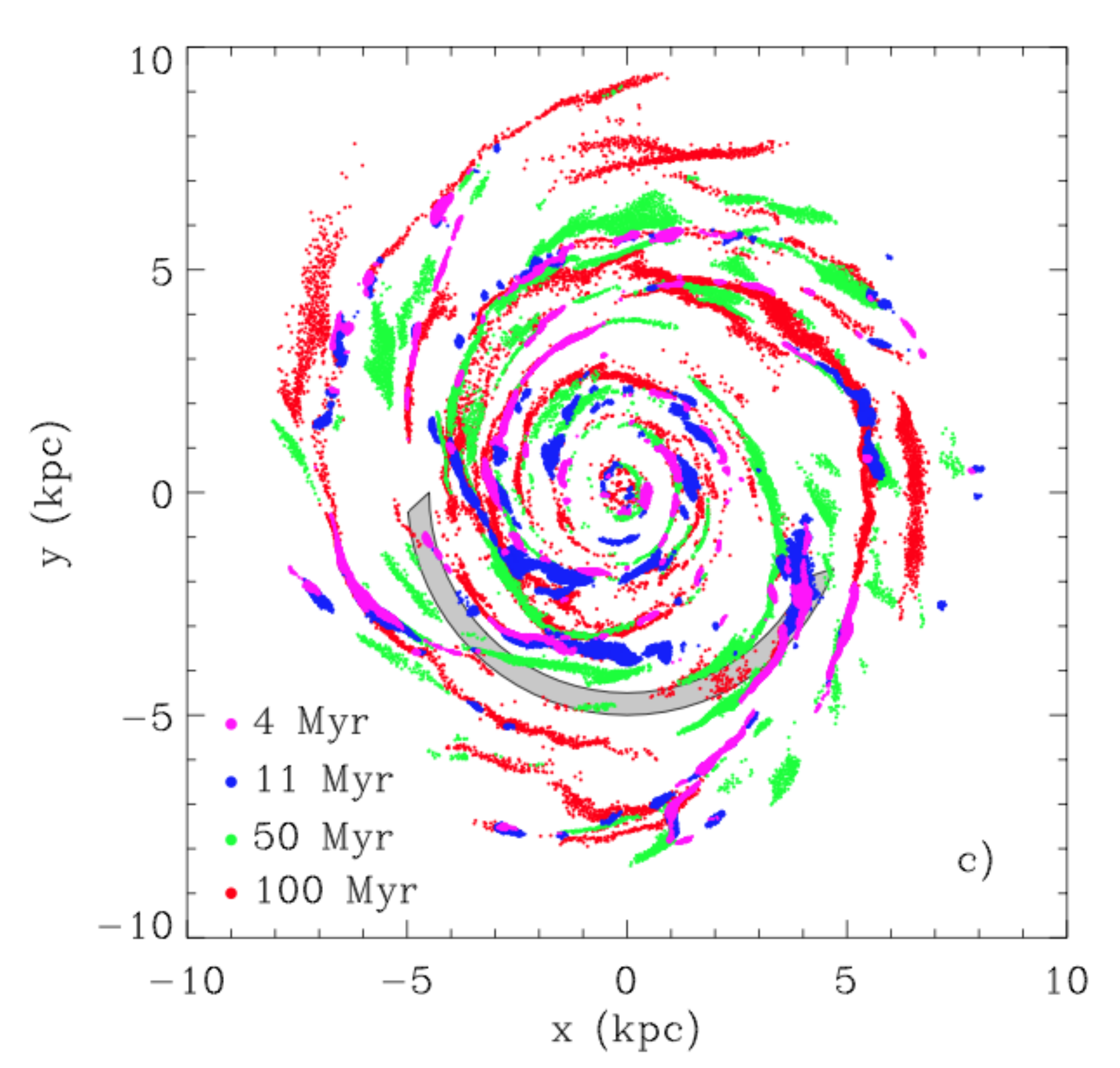}
\includegraphics[scale=0.37]{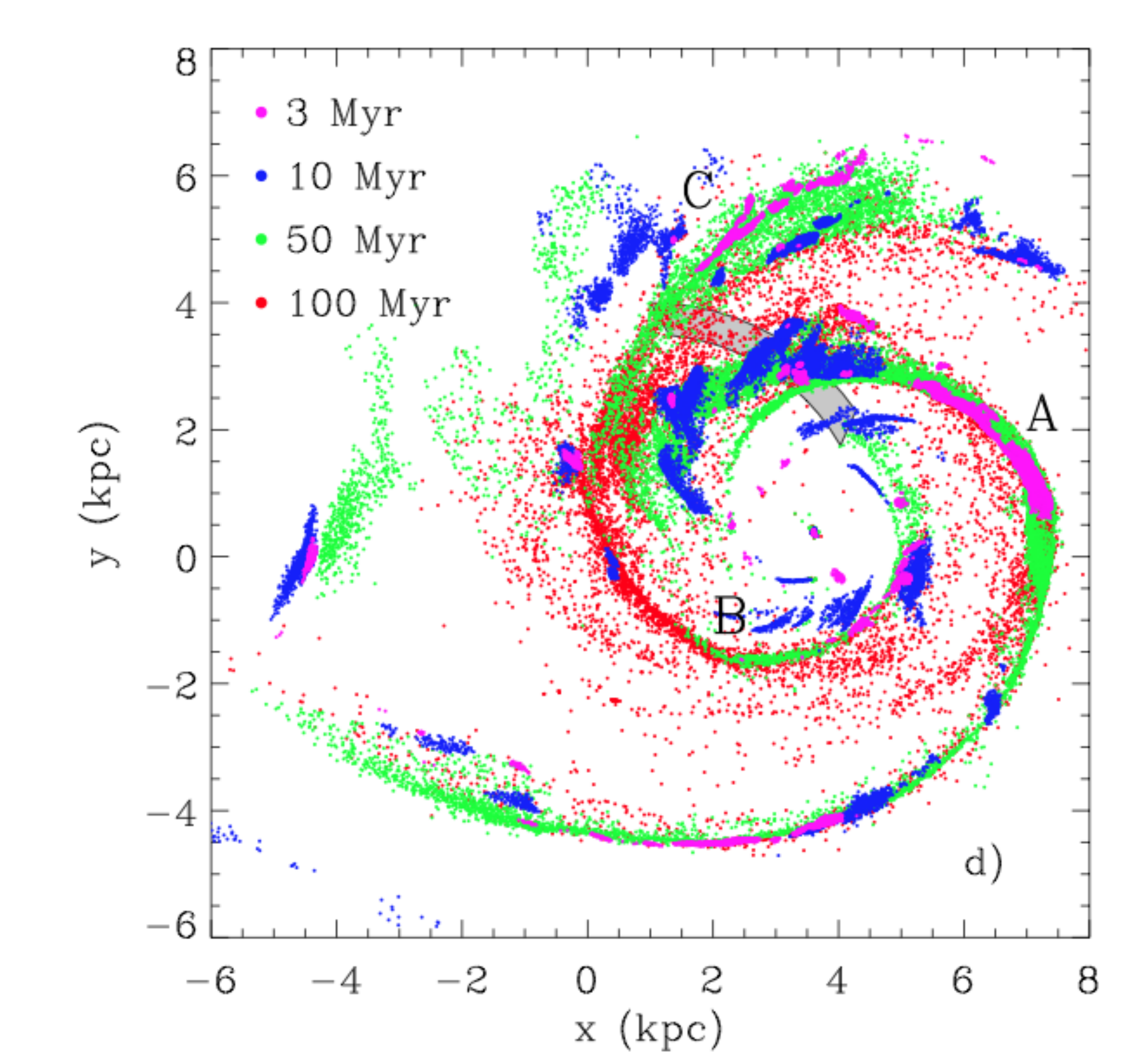}}
\caption{For the galaxy models representing (a) a spiral with a fixed
  pattern speed, (b) a barred galaxy, (c) a flocculent spiral, and (d)
  a tidally induced spiral at the times shown in Fig.~1, we show here
  the estimated positions of star clusters of various ages.  The
  galaxies with a relatively constant pattern speed (the fixed spiral
  (a) and the bar (b)) contain younger stars in the spiral arms or
  bar, with older (100~Myr) stars downstream in the interarm
  regions. The distribution of stellar clusters is more complicated in
  the flocculent (c) and tidally induced spirals (d). For the
  flocculent galaxy, each segment of a spiral arm tends to contain
  clusters of a similar age. In contrast, the tidally induced spiral
  generally shows a complex and somewhat incoherent distribution. The
  grey regions show sections across spiral arms which are used to
  produce the 1D plots showing the distribution of clusters of a given
  age versus distance across the arm (measured as an angle $\theta$)
  and shown in Fig.~4. These are discussed in Section~\ref{xsection}.}
\end{figure*}

\begin{figure}
\centerline{
\includegraphics[scale=0.48]{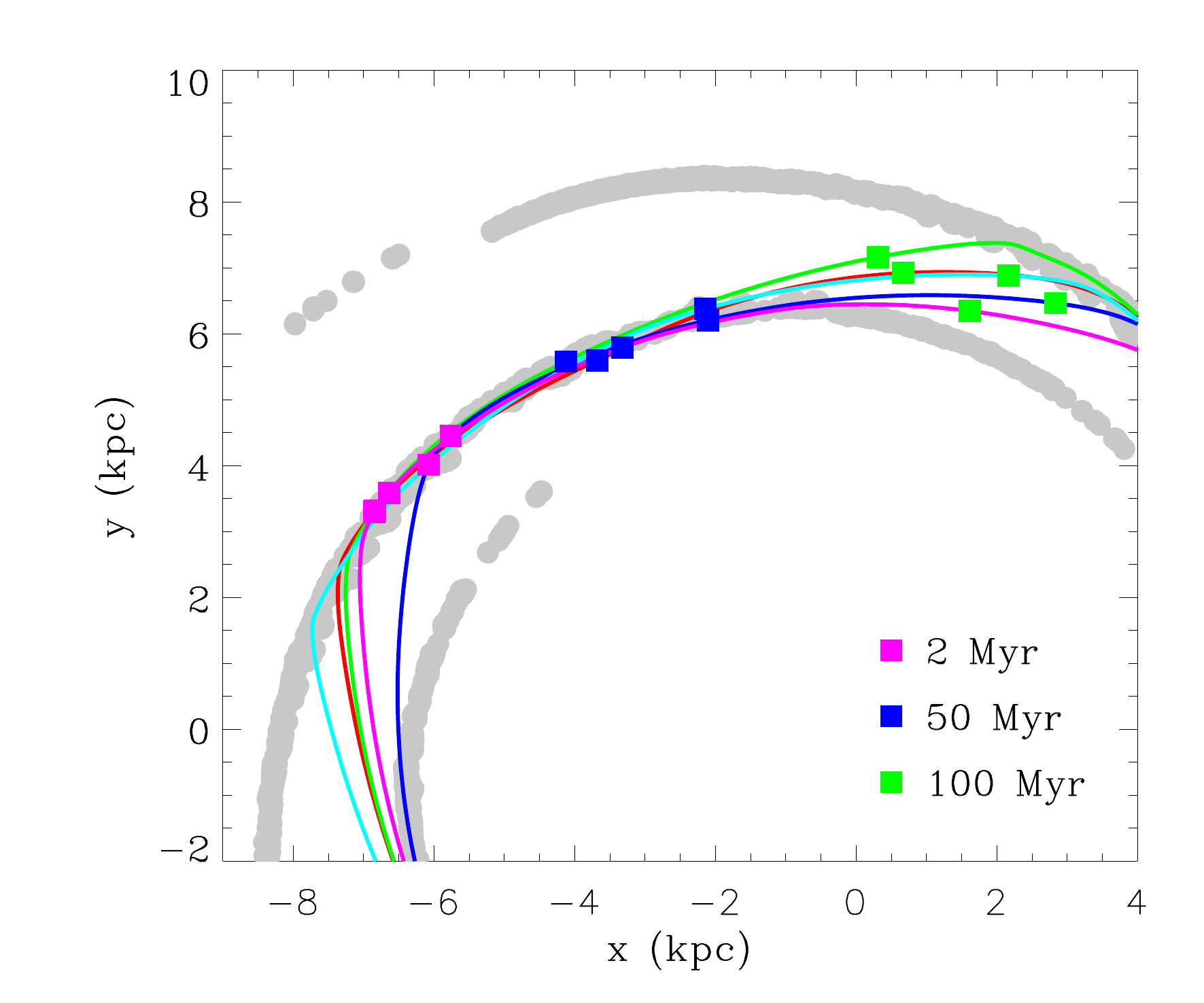}}
\caption{Shown here are the paths of various representative particles
  as they flow through a spiral arm for the case of a galaxy with an
  imposed fixed spiral potential (Fig.~2a), in the rotating frame of
  the potential (the grey shows the spiral pattern). If the flow in
  the rotating frame were exactly steady, these paths would represent
  streamlines of the flow. On each path, the squares indicate where
  gas lies 2~Myr, 50~Myr and 100~Myr after the gas has passed through
  the spiral shock (taken to be the density maximum along the
  path). It can be seen that in this model the gas flows along the
  spiral arm for a time of order 50~Myr, and that the gas has flowed
  into the interarm region by a time of 100~Myr.}
\end{figure}

\begin{figure*}
\centerline{
\includegraphics[scale=0.48]{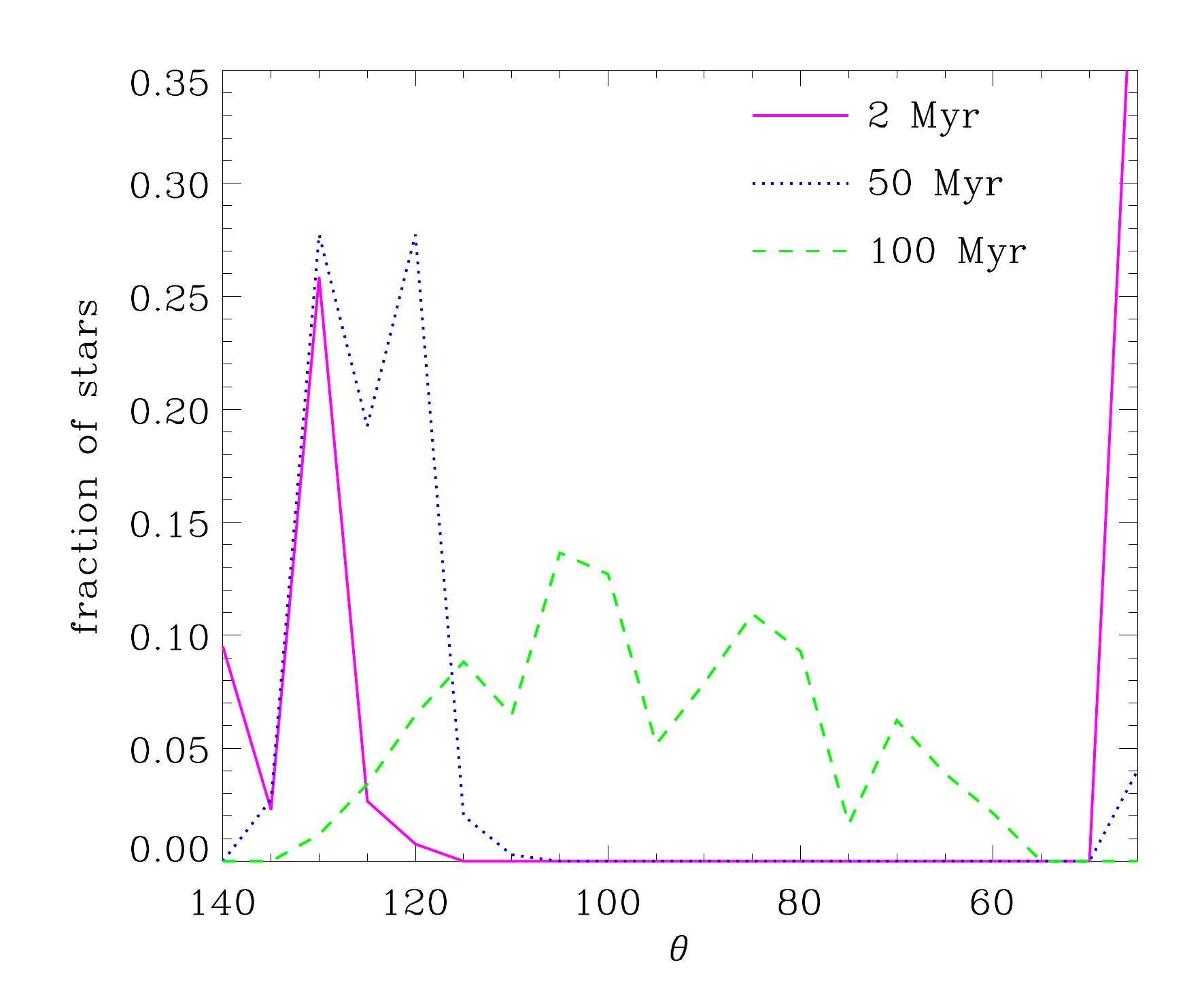}
\includegraphics[scale=0.48]{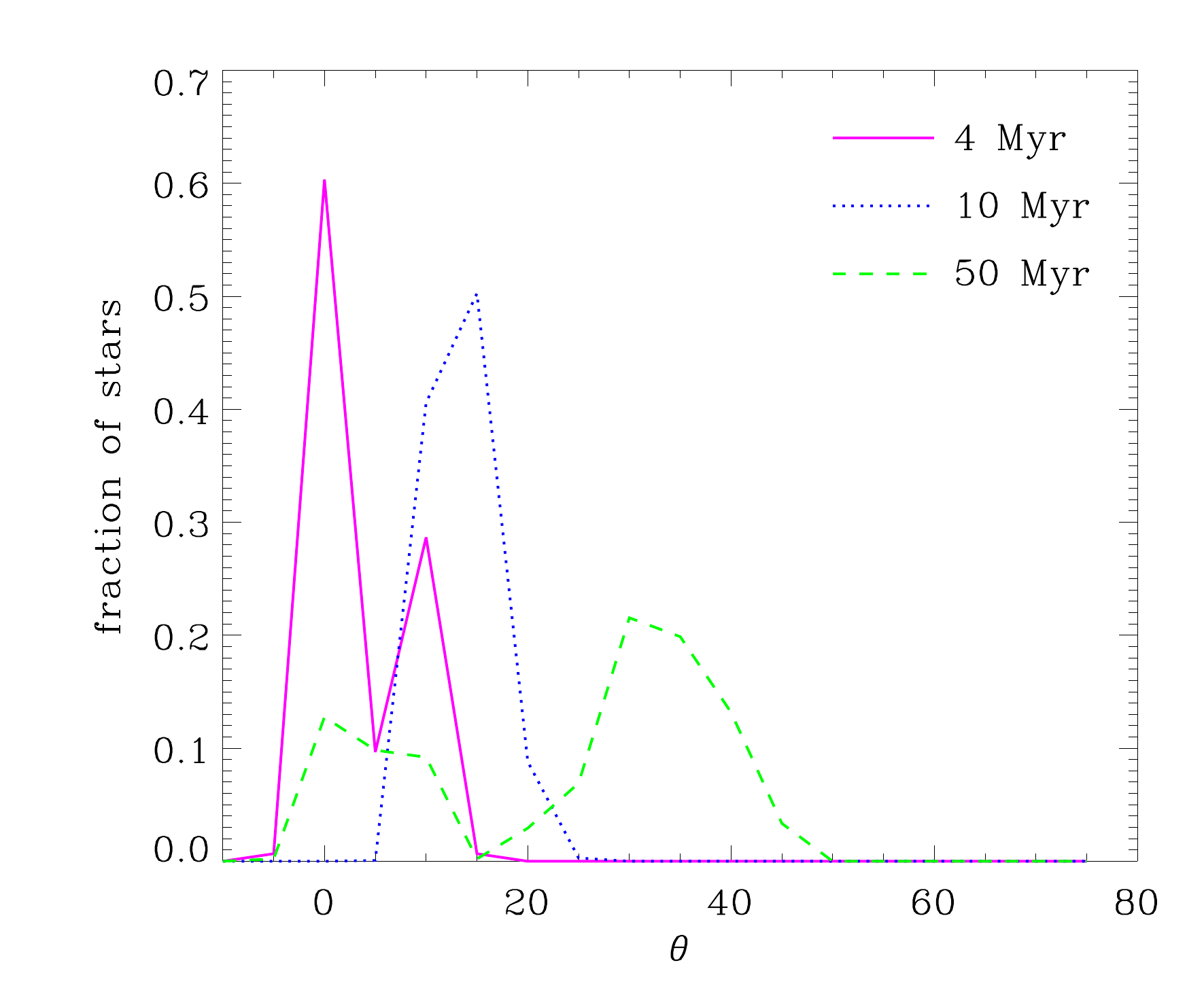}}
\centerline{
\includegraphics[scale=0.48]{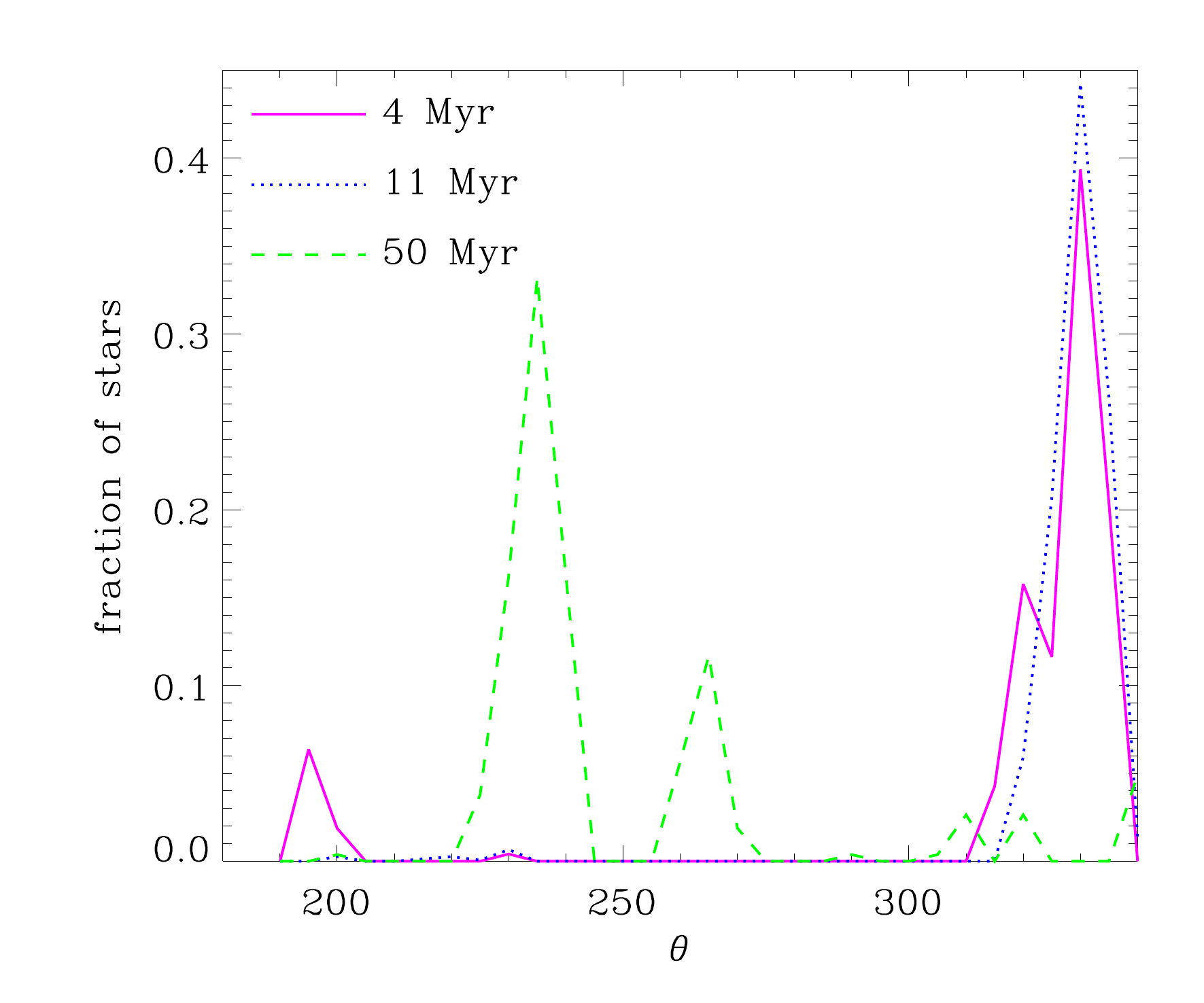}
\includegraphics[scale=0.48]{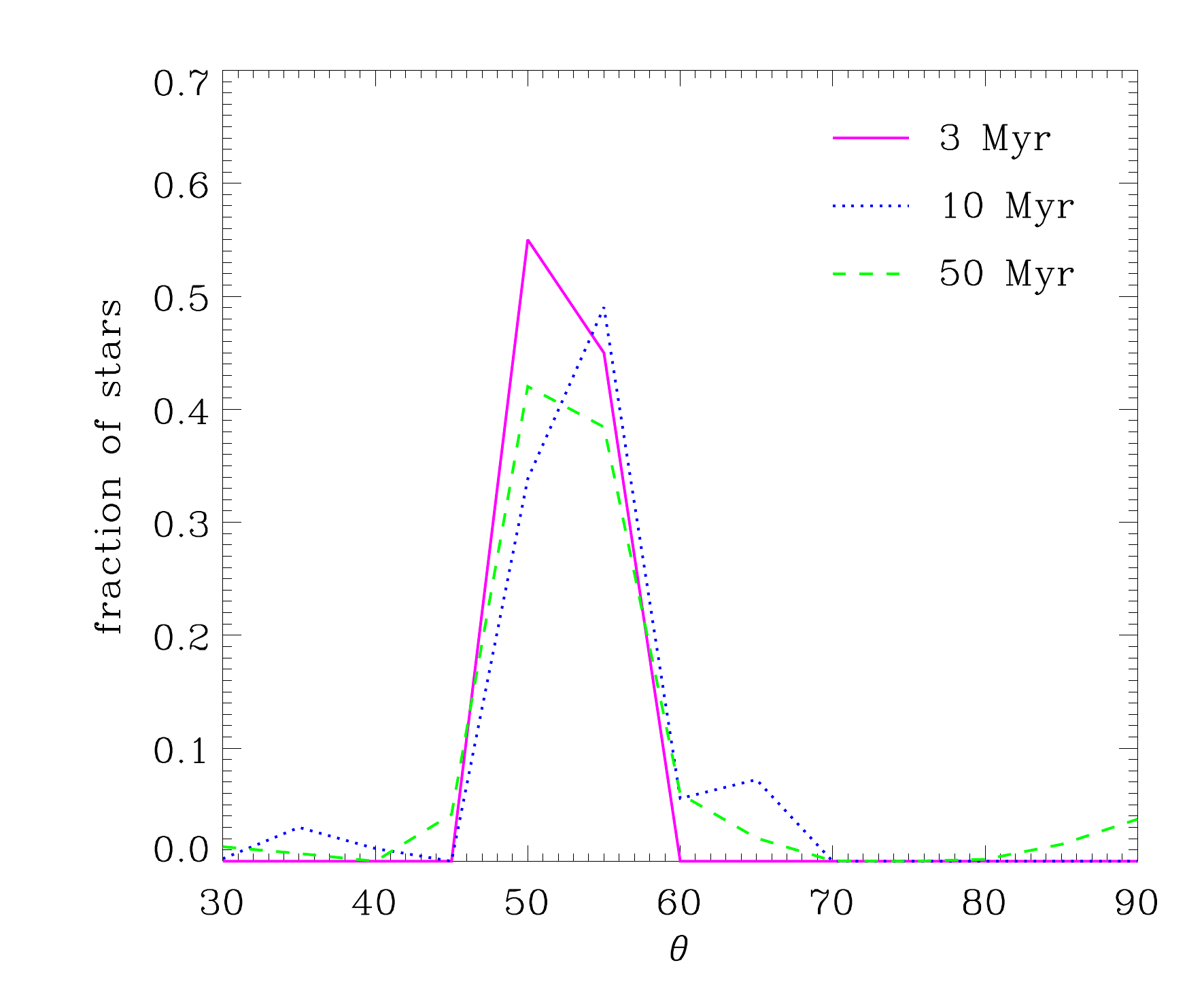}}
\caption{For each galaxy model, corresponding to each spiral
  excitation mechanism, the number of clusters of a given age is
  plotted against distance across a spiral arm. This distance is given
  by the angle $\theta$, where $\theta$ is the angle round the galaxy
  in the radial segment shown in grey in Fig.~2. In all cases $\theta$
  is chosen so that gas flow is from left to right in this
  figure. Both the fixed spiral (top left) and bar (top right) show a
  transition in ages, from youngest to oldest, with the direction of
  rotation (left to right in these plots). The flocculent spiral
  (lower left) shows isolated peaks, each corresponding to cluster of
  a particular age, indicating that spiral arms tend to contain stars
  of similar ages. For the tidally induced spiral (lower right),
  clusters of different ages appear simultaneously in the same region
  of space. As discussed in Section~\ref{m51} the cluster distribution
  is complex in this case, caused by the complicated dynamics inherent
  in the tidal interaction. }
\end{figure*}

\section{Sections across a spiral arm}
\label{xsection}

It is evident that from the point of view of observations of cluster
ages, the main distinguishing feature between the various models of
spiral structure excitation is the way in which cluster ages change
across a spiral arm, or across a spiral arm segment (Fig.~2).

To show this more clearly we plot in Figure~4 the distribution of
clusters of different ages across a particular spiral arm (or arms),
for each of the spiral galaxy models (Sections~\ref{linshu},
\ref{flocculent} and \ref{m51}), and across the bar for the barred
galaxy (Section~\ref{bar}). To do this we choose regions of each
galaxy in the form of circular arcs. These regions are shown as
regions of grey in Fig.~2. \footnote{Thus we produce similar plots to
those shown in \citet{Mart2009}, though they plot a photometric index
as a measure of age, whereas we show the number of clusters of a given
age.}  Except for the tidally induced spiral (Section~\ref{m51}) we
simply took an arc centred on the origin (the centre of the galaxy),
of width 0.5~kpc for the spiral galaxies and 1~kpc for the barred
galaxy. For the tidally induced spiral, an arc is chosen which is more
perpendicular to the spiral arm. In each plot, and for each cluster
age, we normalise the number of clusters to the total number of
clusters of that age. We do not plot the oldest clusters in each case,
as often these have reached the next spiral arm (or bar) and, as
discussed in Section~\ref{ages} at such late times the estimated 
positions of the star clusters are not well modelled.

For each panel in Fig.~4, $\theta$ (the angle round each arc shown in
Fig.~2) is plotted so that the direction of rotation is
anti-clockwise clockwise in Fig.~2. This is in the direction of gas
flow and thus in standard density wave theory we might expect to 
see in Fig.~4 a time sequence of
cluster ages from left to right. However we only observe such a
sequence in the two simulations with a (more or less) constant pattern
speed, which are the spiral with a fixed potential
(Section~\ref{linshu}) and the barred galaxy
(Section~\ref{bar}). These are shown on the top two panels of
Fig.~4. Here we see the highest peaks for the youngest (2~Myr or 4~Myr)
clusters on the left, whilst the 10~Myr and 50~Myr age clusters are
located at successively further to the right.

For the flocculent galaxy (Section~\ref{flocculent}, lower left), we
instead see individual peaks for arms with young clusters
($\theta=195^o$ and $330^o$) or older clusters ($\theta=230^o$ and
$265^o$), but no trend in age across a particular arm.

For the tidally induced spiral (Section~\ref{m51}) we see no clear
pattern at all. The different age clusters are largely coincident,
even though we chose an arc more perpendicular to the arm compared to
the other cases. In particular, we do not see a systematic trend in
ages across an arm. We attribute the difficulty in picking out a clear
trend, such as is seen in the fixed pattern speed and the barred
galaxy, to a variety of effects. At large radii, the arms are largely
tidal, which means gas stays in the arms for long periods (as much as
100~Myr). At smaller radii, there is a significant amount of dense
interarm gas, which appears to be undergoing new star formation. Also
at smaller radii, the timescales for flow between the arms are short,
and thus 50~Myr old clusters in one spiral arm may have formed in the
previous arm. Finally, in this case, the perturbing galaxy makes two
close passes and so effectively induces two sets of tidal arms, as it
orbits the main galaxy. The kink at the point marked A in Fig.~2d
corresponds to one point where two such arms interact and merge.

Overall in a tidal interaction, the stress, shear and compression to
which the spiral arms are subject are likely to vary both in time and
space (and thus the star formation also varies in similar fashion).
This is in contrast to the fixed spiral and the bar, where the forces
acting along the spiral or bar, are more constant in space and time
(and thus in those cases star formation is more uniform along the
spiral arm or bar). 

\section{Discussion and Conclusions}
\label{discussion}

New methods for determining ages of clusters in external galaxies are
now becoming available. We have presented simplified calculations for
the gas flow in spiral galaxies, including a barred galaxy, in which
the spiral structure is induced in the galaxy by different
mechanisms. Using the gas flow so derived, we have estimated in a
simple fashion the likely locations of stellar clusters aging from a
few Myr to around 130 Myr old.  These calculations demonstrate that
different mechanisms for the excitation and maintenance of spiral arms
result in different predictions for the distribution of clusters of
different ages.

If the galaxy has a quasi-static spiral pattern, either caused by
standard density wave theory or by a bar, we expect to see a monotonic
sequence of ages across the spiral arm (or bar) from youngest to
oldest. If the spiral is tidally induced, we would not necessarily
expect to see a clear trend. Rather the distribution of cluster of
various ages depends intimately on the details of the tidal
interaction. In our model, based on M51, the galaxy undergoes
  a double interaction, so the distribution of clusters may be more
  disordered than for a single, simple encounter.
If the galaxy exhibits a predominantly flocculent spiral
structure, induced by local gravitational instabilities, we would
expect individual spiral arm segments to mainly consist of stars of
the same age.

One caveat to our results is that we have determined the expected age
of the clusters from the trajectories of the gas particles, rather
than star particles. However from a calculation ran without gas
pressure, there appears to be little difference between the expected
locations with or without gas pressure, at least until the next spiral
arm passage. A larger uncertainty is that the older the cluster, the
less obvious it becomes which spiral arm the cluster formed in. This
is most problematic for the flocculent galaxy, where spiral arms tend
to collide on timescales of order 100~Myrs, and in the centres of
galaxies.

So far, attempts to find trends in the ages of stars in external
galaxies have mainly used colour gradients. For the traditional view
of density wave triggering of star formation, we would expect to see a
steep gradient on the trailing side of the arm (where there is a
deficiency of young stars), and a much shallower gradient on the
leading side, where young stars have emerged from the spiral arms and
are entering the interarm region.

The analysis of colour gradients in spiral galaxies has produced few
results in agreement with the predictions of density wave theory.  One
of the few positive results is \citet{Gonzalez1996}, who find an age
transition in M99. However \citet{Mart2009} analyse a sample of 13
spiral galaxies and find that in many galaxies, the colour gradients
are opposite to the predictions of density wave theory (i.e. steeper
on the leading side). Furthermore, they often find that for a given
galaxy, one spiral arm follows the predictions whilst the other does
not (e.g. NGC 4254).
 
The results from our simulations show that in a tidally induced, or in
a flocculent spiral, a clear transition in ages is not expected. Thus
the difficulty of finding a transition may simply reflect that those
galaxies do not exhibit quasi-stationary density waves, but rather
that their dynamics are more complex. Alternatively the uncertainties
may simply be too large to properly measure age differences, due to
difficulties correcting for dust and H\textrm{II}, and interarm star
formation. The errors in their data are a sizable fraction of the
gradients they show. Thus age-dating techniques
(e.g. \citealt{Fall2009,Bastian2009}) to directly measure the 
ages of stars may be a better, and more
quantitative, way to differentiate between the theories of spiral
arms. In fact, \citet{Kaleida2010} have already produced a map of the 
spatial distribution of clusters in
 M51. Although they do not show a large enough number of
 clusters for their results to be very conclusive, the lack of any
 pattern in the cluster ages is consistent with results we obtain for our model
 of M51.

According to our results, we would expect to find a clearer trend for
barred galaxies. \citet{Zurita2008} investigate star formation in the
barred galaxy NGC 1530, and do generally find a steeper gradient on
the leading side.  \citet{Popping2010} also investigate ages of
clusters in the barred galaxy NGC 2903, and find the distribution of
stellar ages is much more chaotic, but this is not surprising since
this galaxy is effectively a superposition of a bar with a flocculent
spiral.

\section*{Acknowledgments}

We are grateful to the referee for their helpful comments and suggestions. 
JEP thanks STScI for continued support from their Visitors' Program,
and thanks Ron Allen, Rupali Chandar and Brad Whitmore for valuable
discussions. We also thank Nate Bastian, Eva Schinnerer and Jerry Sellwood for useful comments.
The calculations reported here were performed using 
the University of Exeter's SGI Altix ICE 8200 supercomputer, the 
HLRB-II: SGI Altix 4700 supercomputer at the Leibniz supercomputer
centre, Garching and the SGI Altix at the LMU. The images in Fig.~1
were produced using SPLASH \citep{splash2007}, a visualisation tool for SPH that is
publicly available at http://www.astro.ex.ac.uk/people/dprice/splash.

\bibliographystyle{mn2e}
\bibliography{Dobbs}
\bsp
\label{lastpage}
\end{document}